\documentclass[prla,a4paper,superscriptaddress,twocolumn,showpacs,amsmath,amssymb,floatfix,amstex]{revtex4}

\usepackage[dvips]{graphicx}
\usepackage{bm,pstricks,longtable}

\input{epsf}
\newcommand{\ket}[1]{| #1\rangle}                       %

\newcommand{\bra}[1]{\langle #1}                        %

\begin{document}

\title{Shor's factorization algorithm with a single control qubit 
and imperfections}

\author{Ignacio Garc\'ia-Mata, Klaus M. Frahm and 
Dima L. Shepelyansky}


\affiliation{\mbox{Laboratoire de Physique Th\'eorique - IRSAMC, UPS \& CNRS, 
Universit\'e de Toulouse, 31062 Toulouse, France}}


\date{September 25, 2008}

\begin{abstract}
We formulate and numerically simulate the single control qubit 
Shor algorithm for the case of static imperfections induced by residual 
couplings between qubits. This allows us to study the accuracy 
of Shor's algorithm with respect to these imperfections 
using numerical simulations
of realistic quantum computations with up to $n_q=18$ computational qubits 
allowing to factor numbers up to $N=205193$. 
We confirm that the algorithm remains
operational up to a critical coupling strength $\epsilon_c$
which drops only polynomially with $\log_2 N$.
The obtained numerical dependence of $\epsilon_c$ on $\log_2 N$
is in a good agreement with the analytical estimates that
allows to obtain the scaling for functionality of Shor's
algorithm on realistic quantum computers
with a large number of qubits.
\end{abstract}


\maketitle

\section{Introduction}
\label{sec1}

Shor's factorization algorithm \cite{shor1994}
demonstrates exponential efficiency gain compared to any known classical
algorithm and is definitely the most important quantum algorithm 
in quantum computation \cite{chuang}. 
The possibilities of experimental investigations of the algorithm
are rather restricted due to small number of 
experimentally available qubits and moderate accuracy of 
available quantum gates. Thus the maximal number 
factorized experimentally is $N=15$ 
with a 7-qubit NMR-based quantum computer \cite{chuang1}.

In view of these experimental restrictions the numerical simulations
of Shor's algorithm in presence of realistic imperfections
becomes essentially the only tool for determination of
the conditions of algorithm operability with few tens of
realistic qubits. The first steps have been done in
\cite{cirac1995,paz1996,paz1997} for factorization of $N=15$.
More recently, larger values of $N$ have been studied
with $N$ up to 33  in \cite{nori}
and $N$ up to 247 in \cite{hollenberg2}.
An interesting approach was used in \cite{hollenberg2}:
the Quantum Fourier Transform (QFT) part of Shor's algorithm
has been performed in a semiclassical way using the one qubit
control trick 
(see e.g. \cite{griffiths1996,mosca1999,zalka,plenio,beauregard})
while the modular multiplication
has been performed with up to 20 qubits including
the workspace using the circuit described in 
\cite{hollenberg3}. These works analyzed the effects of 
dynamical phase errors \cite{nori} and discrete qubit flip
errors \cite{hollenberg2,hollenberg3}. 
Another important type of errors
is related to static imperfections induced by 
residual coupling between qubits
which under certain conditions
can lead to quantum chaos melting of a quantum computer
\cite{georgeot2000}. Such type of static imperfections
generally give a more rapid decay of the fidelity
of quantum computation compared to random uncorrelated
phase errors in quantum gates (see \cite{frahm2004} and
Refs. therein). In our recent work 
\cite{garcia2007} we studied the effects static imperfections
for Shor's algorithm
factorizing numbers 
up $N =943$ using up to $L=30$ qubits. 

In this work we combine the two approaches used in 
\cite{hollenberg2,hollenberg3} and \cite{garcia2007}
using Shor's algorithm with  a single control qubit.
This algorithm was introduced and analyzed in
\cite{griffiths1996,mosca1999,zalka,plenio,beauregard}.
It allows us to perform extensive numerical studies
of the effects of static imperfections for Shor's
algorithm factorizing numbers up to 
$N=205193$ that is significantly larger compared to
\cite{hollenberg2,hollenberg3} and \cite{garcia2007}.
Thus, while in \cite{garcia2007} we were able to consider $n_q=10$ 
computational qubits (plus $n_l=2n_q=20$ control qubits
with the total number of qubits $L=30$), we use in the
present  work 
values  up to $n_q=18$ computational qubits. Together with the single 
control qubit this requires a simulation of a quantum algorithm 
with the total number of qubits $L= 19$. 
We remind that without the one control qubit simplification this would require 
$L=54$ qubits that corresponds (if simulated on a classical computer) to an 
array of $2^{54}$ complex elements of $2^{58}$ bytes ($=2^{28}$ GB) in total 
(counting $16$ bytes per complex double precision number). Therefore the one 
control qubit simplification is crucial for the numerical simulation
and allows to increase the value of $N$ by a factor 200
compared to \cite{garcia2007} and $10^3$ compared to
\cite{hollenberg2}.
This allows us to determine the parametric dependence
of the accuracy on the imperfection strength, number of qubits
and number of gates.  
For this we use a simplified but generic model
of imperfections which can be applied to various 
implementations of Shor's algorithm discussed in the literature 
\cite{paz1996,vedral,beckman,zalka,gossett,beauregard,draper,meter,zalka1}.

The paper is structured as follows: in Section~\ref{sec2} we remind the 
standard Shor algorithm (with or without imperfections) and we explain how 
it is possible to obtain a modified version with a single control qubit. 
The results of the numerical simulations are presented in 
Section~\ref{sec3} and the discussion and conclusion are
given in Section~\ref{sec4}. In appendix 
\ref{appendix} we describe the numerical extrapolation scheme, used 
in section \ref{sec3},  allowing to 
determine in an efficient way the inverse participation ratio for an 
unknown random discrete distribution. 


\section{Shor's Alogrithm with one control qubit}
\label{sec2}

Let $N$ be a large integer number of which we want to determine its prime 
factors and $x\ge 2$ a small integer number relatively prime to $N$. The aim 
of Shor's algorithm \cite{shor1994} is to determine the period $r$ defined 
as the minimal positive integer $r$ such $x^r=1$ (modulo $N$). 
As pointed out in \cite{shor1994} the knowledge of the period allows 
(with a certain probability) to obtain a 
non-trivial factor of $N$. The determination of $r$ can be efficiently 
done by a quantum algorithm with $n_q$ computational qubits, chosen such 
that $N<2^{n_q}$, and $n_l=2\,n_q$ control qubits. Actually the 
effective number of $n_q$ may be larger in order to take eventual workspace 
qubits into account which could be needed to realize explicitly the quantum 
modular multiplication in the computational register in terms of elementary 
one- and two-qubit gates. The actual number of 
quantum gates scales like 
$\sim(\log_2 N)^3$, while for  
all known classical computation 
algorithms the scaling is almost exponential.
However, here we do not 
enter into these details and as in Ref.~\cite{garcia2007} we simply assume 
that we can perform this modular multiplication operator in some global way 
not to be specified in the numerical simulation of the quantum computation.
This assumes that there are no quantum errors in the register 
with  $n_l$ control qubits.

In order to keep the following notations simple we associate 
to general operator 
products an ordering from left to right:
\begin{equation}
\label{eq_oporder}
\prod_{j=0}^{n-1} O_j \,\ket{\psi}=O_0\,O_1\,\ldots\,O_{n-1}\,\ket{\psi}\ .
\end{equation}
This convention is necessary to keep the following description unique and 
mathematically precise. 

\subsection{Standard Shor algorithm with $n_l$ control qubits}
\label{subsec_a}

First, we remind the standard Shor algorithm with imperfections 
as modeled in \cite{garcia2007}. 
We start with the initial state
\begin{equation}
\label{eq_initialstate}
\ket{\psi_0}=\ket{0}_{n_l}\,\ket{1}_{n_q}
\end{equation}
and then compute
\begin{eqnarray}
\label{eq_shor1}
\ket{\psi_1} &=& \prod_{j=0}^{n_l-1} H_j \,\ket{\psi_0}\\
\label{eq_shor2}
\ket{\psi_2} &=& \prod_{j=0}^{n_l-1} \left\{
e^{i\delta{\cal H}_j}\,U^{(j)}_{\rm Cmult}\left(x^{2^j} \!\!\!\!\!\mod\! 
N\right)\right\}\,\ket{\psi_1}\\
\label{eq_shor3}
\ket{\psi_3} &=& U_{\rm QFT}\,\ket{\psi_2}\\
\label{eq_shor4}
U_{\rm QFT} &=& R\,\prod_{j=0}^{n_l-1} \left\{H_j
\prod_{k=j+1}^{n_l-1} B_{jk}^{(2)}(\pi 2^{j-k})\right\}\ .
\end{eqnarray}
Here we apply the ordering convention (\ref{eq_oporder}), 
$H_j$ denotes the Hadamard gate acting on the $j$-th control qubit, 
$B_{jk}^{(2)}(\varphi)$ is the controlled two-qubit phase shift gate and for 
later use we also note the simple one-qubit phase shift gate as
$B_{j}^{(1)}(\varphi)$. The operator $R$ reverses the order of the 
$n_l$ control qubits (see ref. \cite{frahm2004}, section 3, 
for more notation details). Here, $U_{\rm QFT}$ is the standard quantum 
Fourier transform \cite{chuang}. 

The operator $U^{(j)}_{\rm Cmult}(x)$ in Eq.~(\ref{eq_shor2})
is the controlled modular multiplication operator 
acting on the computational register as
\begin{equation}
\label{eq_quantmultdef}
U^{(j)}_{\rm Cmult}(x)\ket{y}\equiv\left\{
\begin{array}{ll}
\ket{(yx) \!\!\!\mod N} &\ ,\ y=0,\ldots N-1 \\
\ket{y} &  \ ,\  y=N,\ldots,2^{n_q}-1\\
\end{array}\right.
\end{equation}
if the $j$-th control qubit is $\ket{1}$ and 
$U^{(j)}_{\rm Cmult}(x)\ket{y}=\ket{y}$ if the $j$-th 
control qubit is $\ket{0}$. The operator $e^{i\delta{\cal H}_j}$ denotes the 
error operator which only acts on the computational register (see 
Ref.~\cite{garcia2007} and the next section for details). 
The case of the standard {\em pure} 
Shor algorithm is simply obtained by putting $\delta{\cal H}_j=0$
that eliminates all errors. 

The final step of Shor's algorithm is a measurement of all control 
qubits, thus destroying $\ket{\psi_3}$, 
and resulting in measured numbers from each control qubit: 
$\alpha_j\in\{0,\,1\}$ which provide the (measured) 
control space coordinate by
$a=\sum_{j=0}^{n_l-1} \alpha_j\,2^j$. In the pure case the probability 
distribution of $a$ 
is given by
\begin{equation}
\label{eq:probresult}
P(a)=\frac{1}{Q^2}\sum_{k=0}^{r-1}
\frac{\sin^2(M_k\pi a r/Q)}{\sin^2(\pi a r/Q)} \; .
\end{equation}
where $Q=2^{n_l}$ and $M_k=[(Q-k-1)/r]+1\approx Q/r\gg 1$ only depends 
weakly on $k$.  The function $P(a)$
is composed of $r$ well localized peaks at $mQ/r$ with 
$Q=2^{n_l}$ and $m=0,\,1,\,\ldots,\,r-1$. Since the measured value of 
$a$ is very close to one of these peaks one obtains by a continuous 
fraction expansion 
the value of $r$ provided that $m$ and $r$ are relatively prime (see Refs. 
\cite{shor1994,garcia2007} for more details). 
This algorithm only works with a certain probability since $m$ and $r$ 
may have a common non-trivial factor or because in some rare cases even the 
knowledge of the period $r$ is not sufficient to obtain a non-trivial 
factor of $N$ \cite{shor1994}. In the case 
of imperfections ($\delta{\cal H}_j\neq 0$) the peaks of the probability 
distribution of the control space coordinate become larger and further 
delocalized secondary peaks appear. These effects of imperfections reduce 
furthermore the success probability of Shor's algorithm and can 
be characterized by the inverse participation ratio which is the key quantity 
investigated in \cite{garcia2007} and in section \ref{sec3} of this work. 

\subsection{Reduction to a single control qubit}
\label{subsec_b}

The algorithm described in Eqs.~(\ref{eq_shor1}-\ref{eq_shor4}) requires 
a large number $n_l$ of control qubits and it is therefore quite difficult 
to implement, both in numerical simulations (on a classical computer) 
or eventually in future 
experimental realizations of quantum computers. For numerical simulations the 
large qubit number is especially costly in terms of memory and computation 
time. 

However, for this particular algorithm it is possible to reduce the number 
of control qubits to one single qubit using a scheme
based on a semiclassical implementation of the QFT pioneered by 
Griffiths et al. \cite{griffiths1996} and which was later applied to the 
pure Shor algorithm by Mosca et al. \cite{mosca1999} and Parker et al. 
\cite{plenio}.

In order to understand this significant simplification 
we note that in the above 
Shor algorithm the operator factors associated  to a 
particular value $j$ in the products commute with the operators on the right 
side associated to $\tilde j>j$. This allows to regroup the operator products 
in Shor's algorithm (\ref{eq_shor1}-\ref{eq_shor4}) as follows:
\begin{equation}
\label{eq_shorregroup1}
\ket{\psi_3}=R\,\prod_{j=0}^{n_l-1} V_j\,\ket{\psi_0}
\end{equation}
with operators $V_j$ defined by:
\begin{eqnarray}
\label{eq_shorregroup2}
V_j &=& H_j\,\left\{\prod_{k=j+1}^{n_l-1} B_{jk}^{(2)}(\pi 2^{j-k})\right\}
\times\\
\nonumber
&& \times
e^{i\delta{\cal H}_j}\,U^{(j)}_{\rm Cmult}\left(x^{2^j} \!\!\!\!\!\mod\! 
N\right)\,H_j\ .
\end{eqnarray}
Furthermore, in Eq.~(\ref{eq_shorregroup1}) 
the $j$-th control qubit is not modified by the later factors 
$V_{\tilde j}$ with $\tilde j<j$ and we can therefore measure it 
immediately after the application of the factor $V_j$ (before application 
of the remaining factors). However, after measuring this $j$-th control qubit, 
we need to replace in the remaining factors $V_{\tilde j}$ (with $\tilde j<j$)
the two-qubit control phase shift gates 
$B_{\tilde j j}^{(2)}(\pi 2^{\tilde j-j})$ 
by simple one-qubit phase shift gates which are classically controlled:
$B_{\tilde j}^{(1)}(\alpha_j\,\pi 2^{\tilde j-j})$ where 
$\alpha_j\in\{0,\,1\}$ is the measurement result of the $j$-th 
control qubit. In this way, we see that the information obtained 
from measuring the control qubit $j$ is fed-back
for use of the later values $\tilde j<j$ and thus
the full algorithm can be done with a single control qubit (with $j=0$). 

\begin{figure*}[ht]
\begin{center}
\includegraphics[width=0.9\textwidth]{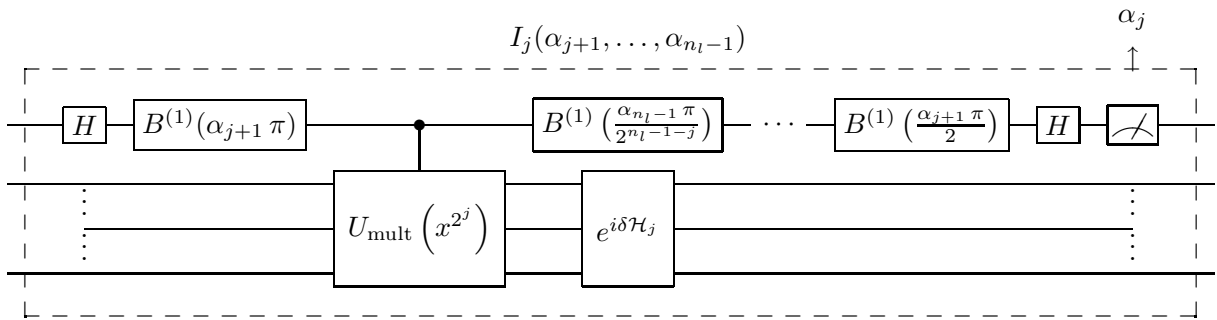} 
\end{center}
\caption{The effective gate $I_j(\alpha_{j+1},\ldots,\alpha_{n_l-1})$ 
which corresponds to one iteration step of Eq. (\ref{eq_onequbit2}). This 
gate includes one measurement producing a classical value
$\alpha_j\in\{0,\,1\}$ and depends on the previously obtained values 
$\alpha_{j+1},\ldots,\alpha_{n_l-1}$. For the initial iteration at $j=n_l-1$ 
we furthermore put $\alpha_{n_l}=0$. We assume that the quantum measurement 
of the single control qubit produces a normalized state obtained from a 
projection and the subsequent normalization. Therefore the effective gate 
conserves the normalization but is not linear and produces 
the classical output $\alpha_j$.
\label{fig1}}
\end{figure*}

\begin{figure}[ht!]
\begin{center}
\includegraphics[width=0.48\textwidth]{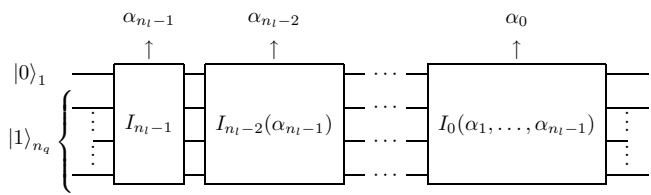} 
\end{center}
\caption{The sequence of effective gates from Fig. \ref{fig1} 
put together providing the one control qubit Shor algorithm and producing 
the classical binary values $\alpha_j$ which allow to reconstruct the 
measured control space coordinate according to Eq. (\ref{eq_reconstruc}).
\label{fig2}}
\end{figure}

Thus, from now on, we assume that the control register 
contains only a single control qubit associated to $j=0$. 
This new algorithm can 
be put in the following recursive form with states $\ket{\varphi_j}$ and 
numbers $\alpha_j\in\{0,\,1\}$ to be determined as:
\begin{equation}
\label{eq_onequbit1}
\ket{\varphi_0}=\ket{0}_1\ket{1}_{n_q}\quad,\quad \alpha_{n_l}=0
\end{equation}
where the first factor in $\ket{\varphi_0}$ refers to the single 
control qubit. Furthermore, for $j=n_l-1,\,n_l-2,\,\ldots,\,1,\,0$ (in this 
order) we compute:
\begin{eqnarray}
\label{eq_onequbit2}
\ket{\tilde\varphi_{n_l-j}} &=& H_0 \,\left\{\prod_{k=j+1}^{n_l-1} 
B_{0}^{(1)}(\alpha_k\,\pi 2^{j-k})\right\}
\times\\
\nonumber
&& \times
e^{i\delta{\cal H}_j}\,U^{(0)}_{\rm Cmult}\left(x^{2^j} \!\!\!\!\!\mod\! 
N\right)
\times\\
\nonumber
&& \times B_0^{(1)}(\alpha_{j+1}\,\pi)\,
H_0\,\ket{\varphi_{n_l-j-1}}\ .
\end{eqnarray}
The state $\ket{\varphi_{n_l-j}}$ is obtained from 
$\ket{\tilde\varphi_{n_l-j}}$ by measuring the single control qubit and 
the measured value will be denoted by $\alpha_j$. Due to the projection of the 
measurement $\ket{\varphi_{n_l-j}}$ has the form:
\begin{equation}
\label{eq_project1}
\ket{\varphi_{n_l-j}}=\ket{\alpha_j}\,\ket{\hat\varphi_{n_l-j}}
\end{equation}
where $\ket{\hat\varphi_{n_l-j}}_{n_q}$ is a state which only lives in the 
computational register. The state (\ref{eq_project1}) 
will be used as the initial state 
in the next step with $j-1$ and since $\alpha_j$ may be 1 the application of the 
Hadamard gate $H_0$ may provide a ``wrong sign'' in this case:
\begin{equation}
\label{eq_project2}
H_0\,\ket{1}\,\ket{\hat\varphi_{n_l-j}}=\frac{1}{\sqrt2}\left(
\ket{0}-\ket{1}\right)\,\ket{\hat\varphi_{n_l-j}}
\end{equation}
and therefore we have introduced in Eq.~(\ref{eq_onequbit2}) the 
additional gate $B_0^{(1)}(\alpha_{j}\,\pi)$ (for $j-1$) such that:
\begin{equation}
\label{eq_project3}
B_0^{(1)}(\alpha_{j}\,\pi)\,H_0\,\ket{\alpha_j}\,
\ket{\hat\varphi_{n_l-j}}_{n_q} =\frac{1}{\sqrt2}\left(
\ket{0}+\ket{1}\right)\,\ket{\hat\varphi_{n_l-j}}
\end{equation}
which is indeed the desired initial condition for the next step. In the 
above algorithm we also introduce artificially $\alpha_j=0$ for $j=n_l$ 
which is 
normally not relevant and simply provides a proper functioning of the 
iteration at the first step at $j=n_l-1$. 
The quantum circuits associated to the iteration 
(\ref{eq_onequbit2}) and the one control qubit Shor algorithm
are shown in Figs.~\ref{fig1} and 
\ref{fig2} respectively.

The final operator $R$, which inverses the order of the control bits, can be done 
classically by the reconstruction of the measured control space coordinate:
\begin{equation}
\label{eq_reconstruc}
a=\sum_{j=0}^{n_l-1} \alpha_{n_l-1-j}\,2^j\ .
\end{equation}
The one control qubit version of Shor's algorithm reproduces exactly the 
same probability distribution of the control space coordinate as 
the standard Shor algorithm described above in section \ref{subsec_a}.

\section{Numerical Results}

\label{sec3}

The probability distribution $P(a)$ in Eq.~(\ref{eq:probresult}) 
only weakly depends on $k$. 
This expression, viewed as a function of a {\em real} variable $a$, 
has $r$ equidistant strongly localized 
peaks of width unity, of height $1/r$ and located at 
$m Q/r$ with $m=0,1,\ldots,r-1$. For integer values of $a$ 
the real peak height is probably smaller than $1/r$ since the exact position 
is not reached. However, the choice $n_l=2n_q$ ensures that there is at most 
only one integer value of $a$ close to the exact peak \cite{shor1994}. 

We model static imperfections generated by residual couplings between qubits
in the frame of the generic quantum computer model analyzed in \cite{georgeot2000}.
These residual  static imperfections produce
 additional unitary rotations $U_s=e^{i\delta H}$ in the quantum gates. 
As in Refs.~\cite{frahm2004,garcia2007} we model the static imperfections 
with the effective perturbation operator
\begin{equation}
\label{heffmodel}
\delta{\cal H}_j=\sum_{i=0}^{n_q-1}\delta_i(j)\sigma_i^{(z)}+2
\sum_{i=0}^{n_q-2}J_i(j)\sigma_i^{(x)}\sigma_{i+1}^{(x)}
\end{equation}
where $\sigma_i^{(\nu)}$ are the Pauli operators acting on the $i$th qubit 
(of the computational register) 
and $\delta_i(j),\,J_i(j)$ are random coefficients distributed according to:
\begin{equation}
\label{Jdelta}
\delta_i(j),\ J_i(j)\in[-\sqrt{3}
\epsilon,\sqrt{3} \epsilon]\ .
\end{equation}
As it was done in \cite{garcia2007}, 
we consider two models where the random coefficients $\delta_i(j),\,J_i(j)$ 
are different for each value of $j$ ({\em generic} imperfection model) 
or equal for all values of $j$ ({\em correlated} imperfection model).

Since, for $\delta{\cal H}_j\neq 0$, the success of the algorithm depends 
essentially on the probability of hitting one of $r$ peaks in the process of 
the $a$-measurement the most direct way  to study this probability is 
by {\em clashing\/} all the peaks into one, 
or in other words, adding them 
all together by taking $a$ modulo $s$ where $s$ is the nearest
integer value of the ratio $Q/r$ and thus reducing all probabilities
inside one cell with $s$ states. 
In this way we obtain a new distribution of global search probability  $W(a)$:
\begin{equation}        
        \label{eq:clash}
W(a)=\sum_{j=0}^{r-1} P([a+s+jQ/r] \mod s)
\end{equation} 
where now $a=-s/2,\ldots,s/2-1$ (the difference of 
$a$ for $P$ and $W$ is clear from the context)
and $s \approx Q/r$ is the distance between peaks.
For the ideal algorithm this global probability
$W(a)$ has one peak at $a=0$ while in the case of imperfections the 
main peak may become larger and secondary peaks appear. We also use the original 
notation of \cite{shor1994} putting $a=c$.

As in Ref.~\cite{garcia2007} we study the ``delocalization'' effects 
of quantum chaos due to 
the imperfections by computing the inverse participation ratio associated 
to the global probability distribution $W(a)$:
\begin{equation}
        \label{eqipr}
\xi=(\sum_a |W(a)|^2)^{-1}
\end{equation}

In Ref.~\cite{garcia2007}, the complete state $\ket{\psi_3}$ was calculated 
from a classical simulation (with up to 30 qubits: $n_l=20$ and $n_q=10$) 
thus allowing to determine exactly all the key quantities such as the full 
probability distributions $P(a)$ and $W(a)$, inverse participation ratio, 
variance and this without actually measuring and 
destroying the state $\ket{\psi_3}$. However, the computation of these 
quantities 
is only possible due to the (quite expensive) {\em classical} 
simulation of a {\em quantum} algorithm. In fact, the original Shor 
algorithm actually contains a measurement of the control space variable 
$a$. This point is rather important since the one control qubit version 
of Shor's algorithm is essentially based
on measurements and gives a significant reduction
of numerical computational efforts. In such a case,
as with a real quantum computer, we are not able 
to determine directly the exact probabilities $W(a)$ 
(unless $\delta{\cal H}_j=0$ where the theoretical formula 
(\ref{eq:probresult}) applies) 
but we may draw as many values $a$ with this probability distribution as we 
want simply by repeating the simulation of Shor's single control qubit 
algorithm with other sequences of measurement outputs. 
Thus, many repetitions of many random results of measurements 
is the prize to pay for the reduction of number of 
control qubits from $n_l$ to one. 

\begin{figure}[ht!]
\begin{center}
\includegraphics[width=8cm]{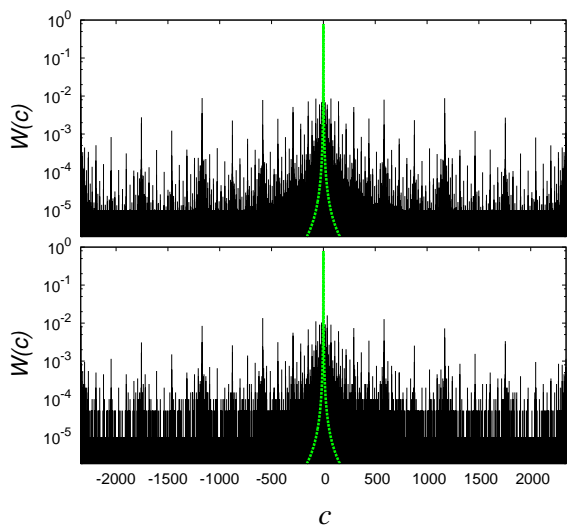} 
\end{center}
\caption{(Color online)
The global probability distribution $W(c)$, $c=a$,
as defined in Eq.~(\ref{eq:clash}), 
averaged over $N_R=10$ realizations of random static imperfections, 
for $N=493$, $x=2$, $r=56$, $\epsilon=0.04$,
disorder realizations are identical for top and bottom panels.
The top panel corresponds to the complete computation of 
Shor's algorithm as in \cite{garcia2007} using in total
$27$ qubits, 
the bottom panel corresponds to the histogram 
computed using $N_{\rm meas} \sim 12\,000$ 
measurements (error tolerance $\sim 2\%$ for 
the associated inverse participation ratio)
in the simulation of Shor's single qubit algorithm
with in total $10$ qubits. 
The grey/green dashed line corresponds to the probability at $\epsilon=0$. 
\label{fig3}}
\end{figure}

Of course we may replace the exact probabilities $W(a)$ by histogram 
probabilities which will be as accurate as we want provided that the number 
of series of measurements is sufficiently large. 
If we want to determine the full distribution 
the classical simulation of Shor's single control qubit algorithm is no 
longer advantageous (in computation time) as compared to the 
direct simulation of the full Shor algorithm as done in \cite{garcia2007}. 
But if we need to know only certain averaged characteristics of the distribution
$W(a)$, e.g. the inverse participation ratio,
then Shor's single control qubit algorithm becomes much more efficient
compared to the approach used in \cite{garcia2007}. To compare 
the validity of these two approaches we verified for 
some small numbers of $n_l$ and $n_q$ that the histogram distribution 
obtained from the simulation of Shor's single control qubit algorithm 
reproduces very accurately  all details of
the exact distribution obtained from a full Shor 
algorithm simulation if both cases are simulated with the identical disorder 
realization for $\delta{\cal H}_j$. In Fig. \ref{fig3} we show as an example 
a comparison of the distribution $W(a)$ obtained from both types of 
simulations and for an average over 10 disorder realizations (identical 
disorder realizations are used for two computational methods). 

The single qubit Shor algorithm is advantageous for the computation 
of the inverse participation ratio $\xi$ provided $\xi$ is not too large 
because this requires less measurement series samples 
than for the full histogram to achieve 
a reasonable accuracy. However, a simple replacement of $W(a)$ 
by the histogram probabilities in (\ref{eqipr}) is not very optimal for 
modest sample numbers $N_R$ since 
the average of $\xi^{-1}$ (with $\xi$ obtained from histogram probabilities) 
scales with the sample number and is not identical with the exact value
of $\xi^{-1}$ (with $\xi$ obtained from the exact probabilities $W(a)$). 
In appendix \ref{appendix} we describe a numerical extrapolation scheme 
that allows to take this 
into account and to determine a more accurate value of
$\xi$ for a finite sample number and also to control its statistical variance. 
The numerical results of $\xi$ presented in the following have been obtained 
by this extrapolation scheme and choosing a sample number to achieve 
a 2\% precision. 

\begin{figure}[ht!]
\begin{center}
\includegraphics[width=8cm]{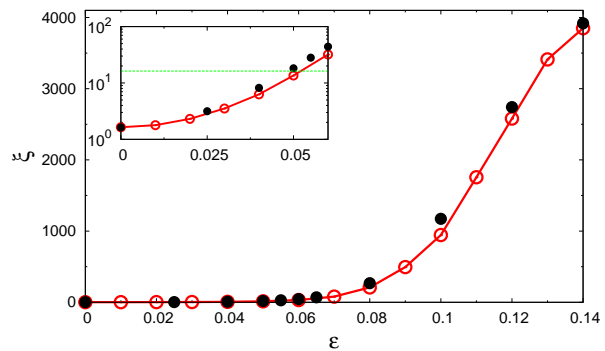} 
\end{center}
\caption{(Color online) Averaged IPR $\xi$ given by (\ref{eqipr})
as a function of $\epsilon$
for $N=493$, $x=2$. The full circles symbol correspond to the simulation 
of the full Shor algorithm used in \cite{garcia2007} 
with the complete control register, 
therefore $n_{\rm tot}=27$. The open circles correspond to the IPR
obtained by the single qubit Shor algorithm 
with the total number of qubits $L=n_q+1=10$
and 
the extrapolation scheme with an error tolerance of 2\% that required
about $N_{\rm meas} \sim 12000$ measurements.  
The inset shows 
the dependence for small $\epsilon$ values, 
the horizontal dashed line marks the 
quantum chaos border defined by the condition
$\xi(\epsilon_c)=10\, \xi(0)$.
There are $N_R=10$ disorder realizations which are identical for
full and open circles.
\label{fig:ipr}
\label{fig4}}
\end{figure}

In Fig. \ref{fig4} we show the dependence of $\xi$ 
on the strength of imperfections
$\epsilon$.  The values of $\xi$ 
are obtained by the full simulation of Shor's algorithm as 
in \cite{garcia2007} and by the simulation of the single qubit Shor 
algorithm using the extrapolation scheme. We see that both methods 
give very close results even for large values of $\xi$.
Thus we may use the more efficient single qubit algorithm to
test effects of imperfections for factorization of numbers $N$
much larger than those of \cite{garcia2007}. 

\begin{figure}[htb!]
\begin{center}
\includegraphics[width=8cm]{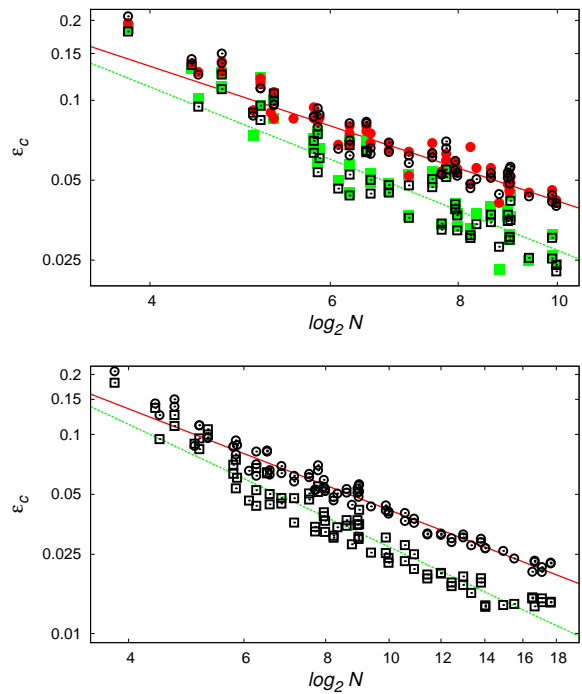} 
\end{center}
\caption{Dependence of the critical imperfections strength
$\epsilon_c$ on $\log_2 N$ in log-log scale,  
$\epsilon_c$ is obtained from  the criterion
$\xi(\epsilon_c) = 10 \xi(\epsilon=0)$, 
where $\xi$ is the inverse participation ratio (\ref{eqipr}).
The top panel shows the results 
obtained in \cite{garcia2007} marked by full symbols
(full, red, circles are for the generic 
imperfection model; 
full, green, squares are for the correlated imperfection model)  
and the results of this work marked by open symbols
(open circles: the generic imperfection model;  
open squares: the correlated imperfection model).
The bottom panel shows the data of this work 
up to $N=205193$ (open symbels).
The straight lines show the fits $\epsilon_c = B/(\log_2 N)^{\beta}$
of open symbols data
for the two models of imperfections
in the interval $5.5 \le \log_2 N \le 18$ with
$B=0.7877 \pm 0.073$, $\beta =1.275 \pm 0.045$ 
for the generic imperfection model (solid, red, line), 
$B=0.958 \pm 0.154$, $\beta =1.546 \pm 0.08$ 
for the correlated imperfection model (dashed, green, line).
 \label{fig:epsc}}
\label{fig5}
\end{figure}

The results of Figs.~\ref{fig3},\ref{fig4} show that with 
the increase of imperfections strength $\epsilon$
the peak in the distribution $W(a)$ is washed out
and the quantum chaos destroys the operability of the algorithm.
As in \cite{garcia2007} the critical strength of the
imperfections $\epsilon_c$ at the quantum chaos border
can be approximately determined by the 
condition $\xi(\epsilon_c) = 10 \xi(\epsilon=0)$.
In Fig. \ref{fig5} we show the dependence of $\epsilon_c$
on the number $N$
factorized by Shor's algorithm in a 
log-log scale. The dependence on $N$
can be described as
\begin{equation}
\epsilon_c=\frac{B}{(\log_2 N)^\beta}
\label{qborder}
\end{equation}
with numerical constants $B$ and $\beta$.
The fit of numerical data done for large $N$ values
in the interval $5.5 \le \log_2 N \le 18$ gives
$B=0.7877 \pm 0.073$, $\beta =1.275 \pm 0.045$ for the generic 
imperfection model and $B=0.958 \pm 0.154$, $\beta =1.546 \pm 0.08$ for the 
correlated imperfection model. 
The values for the exponent $\beta$ differ slightly from those obtained 
in \cite{garcia2007} where we had
$\beta=1.420 \pm 0.054$ for the generic imperfection model
and $\beta=1.523 \pm 0.068$ for the correlated imperfection model.
In view of strong fluctuations
related to the  arithmetic properties of $x,\, r$ and $N$
we can consider that the agreement with the 
results obtained in \cite{garcia2007}
for not very large values of $N < 1000$
is rather good. The new data
allowed to increase the values of $N \leq 205193$ 
by a significant factor 200
that gives more accurate values of the exponent $\beta$.
The obtained values of $\beta$ are close to the values
given by the theoretical estimates \cite{garcia2007}
with $\beta=1$ for the generic imperfection model 
and $\beta=1.5$ for the correlated imperfection model.
We attribute the deviations of numerical values of $\beta$ 
from the theoretical values to strong arithmetical fluctuations
which require a large scale of $\log_2 N$-variation.
We also note that the statistical fluctuations
related to randomness and disorder in 
realizations of imperfections are relatively small
since the standard deviation from disorder average
gives an error bar which is approximately of 
the symbol size in Fig.~\ref{fig5}
(the same is true for the data of \cite{garcia2007}).

To give more information we present the results and parameters of our
numerical simulations in the
Table~\ref{table1} for large $N$ values $10^3 \le N \le 2\cdot 10^5$ 
which were inaccessible in Ref.~\cite{garcia2007}. 
\begin{table}[htb]
\caption{Table for the values of $\epsilon_c$ plotted 
in Fig.~\ref{fig:epsc}, for values of $N$ greater than the
ones of \cite{garcia2007}. In Fig.~\ref{fig:epsc} 
we see that below $N \approx 1000$ the 
values of $\epsilon_c$ quite well coincide  
with the ones computed previously in \cite{garcia2007}. 
For simplicity of the table
we rounded the numbers to three significant digits. 
Number of realizations $N_R$ is approximate. The number of measurements 
$N_{\rm meas}$ gives the order of magnitude 
of the maximum number of measurements
for both imperfections models 
(generic $^{(1)}$ and correlated $^{(2)}$ 
imperfection models). Here $L=n_q+1$ 
is the total number of qubits. 
\label{table} }

\begin{ruledtabular}
\begin{tabular} {lccccccr}  
$N$&$L$& $\epsilon_c^{(1)}$& $\epsilon_c^{(2)}$&$x$&$r$&$N_R$&$N_{\rm meas}$\\
\hline
1007 & 11 &0.04& 0.023 & 4 & 234 &30&  $1\times10^6$ \\
 1517& 12 & 0.037& 0.023& 2 & 90 &15&$ 5\times 10^5$  \\
1517 &12  &0.040& 0.028& 3 & 72 &15& $3\times 10^5$  \\
1927 &12  & 0.036& 0.021& 2 & 460 &15& $1\times 10^5$  \\
1927 &12  & 0.038& 0.025& 3 &184  &15&  $4\times 10^5$ \\
2773 &13  & 0.032& 0.020& 2 & 1334 &15 &$5\times 10^5$   \\
2773 &13  &0.031 &0.019 & 3 & 667 &15 & $6\times 10^5$  \\
4087 &13  & 0.032& 0.020& 2 & 660 &16 & $2\times10^6$  \\
4087 &13  &0.031 & 0.020& 4 & 330 &16 &   $2.5\times10^6$ \\
5609 &14  &0.029& 0.017& 2 & 1365 &15& $1.5\times 10^6$  \\
5609  &14 &0.030& 0.018& 3 & 2730 &15 &$1.5\times 10^6$ \\
8051  &14&0.031& 0.019& 2 &  1968&15 & $2.5\times 10^6$  \\
 8051 &14 &0.030 &0.017 & 4 & 984 &15 & $2\times 10^6$\\
10403 & 15 &0.028&0.016 & 2 & 5100 &6 & $2\times 10^6$  \\
14351 & 15 &0.030&0.019 & 2 & 28 &15& $6\times 10^5$ \\
14351& 15  &0.029&0.018 & 3 & 1008 &15 & $1\times 10^6$ \\
 16631 &16&0.027& 0.014& 2 & 1663 &4& $2.7\times 10^6$  \\
16631 &16 &0.027 &0.014& 3 & 8315 &4 & $2.1\times 10^6$\\
31313 & 16 &0.026 &0.014& 2 & 7740 &15 &$2.1\times 10^6 $   \\
47053 & 17 &0.024 &0.014& 2 & 7770 &3 & $1.8\times 10^6$  \\
 95477 & 18& 0.020&0.015& 2 & 15810 &10 &  $3.7\times 10^5$ \\
104927 & 18& 0.023&0.015  & 2 & 4740 &8&  $6.6\times 10^5$ \\
  141367& 19&0.020 &0.014  & 2 & 23436 &3&  $3.8\times 10^5$  \\
141367 & 19& 0.021&0.015  & 4 & 11718 &3&  $3.8\times 10^5$  \\
 205193 &19& 0.022&0.014  & 2 & 4256 &3&  $5\times 10^5$ \\
 205193& 19&0.022 & 0.014 & 4 & 2128 &4& $5.1\times 10^5$  \\
\end{tabular}
\end{ruledtabular}
\label{table1}
\end{table}

\section{Conclusion}
\label{sec4}
The extensive numerical simulations performed in this
work allowed to analyze the accuracy and operability 
bounds for Shor's algorithm in presence of realistic static 
imperfections. The results show that above the quantum chaos border
$\epsilon_c$ given by Eq.~\ref{qborder} the algorithm becomes not
operational while below the border the factorization
can be performed. This border drops only polynomially
with the logarithm of factorized number $N$.
The algebraic power $\beta$ of this decay is close
to the theoretical estimates obtained in \cite{garcia2007}.
The numerical values of $\beta$ are close to the values
obtained in \cite{garcia2007}
where the factorization was studied for
significantly smaller values of $N$
compared to the present work.
Due to that we think that our results give
the real asymptotic value of the algebraic
exponent $\beta$ for the quantum chaos border in 
Shor's algorithm in presence of static imperfections.
Even if the values of $\beta =1$ or $1.5$ 
are relatively low still the accuracy requirements for quantum
gates become rather restrictive if one wants to 
factorize such large $N$ values as those
used in classical computers 
(see more detailed discussion in \cite{garcia2007}).

This work was supported in part by the EC IST-FET project EuroSQIP.
For numerical simulations we used the codes of Quantware Library \cite{qwlib}.

\appendix

\section{Efficient numerical determination of the IPR for a 
discrete random variable}

\label{appendix}

Let us consider a discrete random variable $x$ with possible values 
$x=0,\,1,\,\ldots,\,Q-1$ and probabilities $p(x)\ge 0$ properly 
normalized $\sum_x p(x)=1$. The inverse participation ratio of $x$ is 
defined as:
\begin{equation}
\label{app_eq1}
\frac1\xi=\sum_x p^2(x)\ .
\end{equation}
Here $\xi$ denotes roughly the number of possible $x$ values with a 
significant (``maximal'') probability. In the case of quantum states 
$\ket{\psi}=\sum_x \psi(x)\,\ket{x}$ with $p(x)=|\psi(x)|^2$ the quantity 
$\xi$ is also refereed as the (inverse participation ratio) localization 
length. Obviously, $\xi$ is easily computed provided the exact probabilities 
$p(x)$ are known. In Ref.~\cite{garcia2007}, this was indeed the case since 
we were able to calculate the full quantum state after application of 
Shor's algorithm but before measuring the control space variable. 
However, in this work, where we use the one-control-qubit version 
of this algorithm as described in section \ref{sec2}, this is no longer 
possible since the exact values of $p(x)$ are not known and we are 
``only'' able to draw an arbitrary number of values $x_j$, 
$j=1,\,\ldots,\,R$ using this probability distribution by simply 
repeating the one-control-qubit Shor algorithm $R$ times. In the limit 
$R\to\infty$ this should in principle allow to recover $p(x)$ and $\xi$ 
with sufficient accuracy but for ``moderate'' values of $R$ (e.~g.: 
$R\approx 10\xi - 100\xi$) this is not very precise and can be improved by 
a kind of extrapolation scheme in $R$ which we will now explain.

Suppose that $x_1,\,\ldots,\,x_R$ are independent random variables with the 
same (unknown) probability distribution $p(x)$. For a given set of 
$x_1,\,\ldots,\,x_R$ (representing ``numerically obtained values'') 
we introduce the histogram probabilities by:
\begin{equation}
\label{app_eq2}
p_R(x)=\frac1R\,n_{\{x_j\}}=\frac1R \sum_{j=1}^R\,\delta_{xx_j}
\end{equation}
where $n_{\{x_j\}}$ is the number of $x_j$ values being equal to $x$. 
Obviously the average of $\delta_{xx_j}$ with respect to $x_j$ is: 
$\langle \delta_{xx_j}\rangle=\sum_{x_j} p(x_j)\,\delta_{xx_j}=p(x)$ 
and therefore $\langle p_R(x)\rangle = p(x)$. For this simple quantity 
the average histogram value indeed coincides with the exact value. 
However, this 
is not the case for other quantities. Let us for example consider the IPR 
value obtained by the histogram probabilities:
\begin{equation}
\label{app_eq3}
\frac1{\xi_R}=\sum_x p_R^2(x)
\end{equation}
with the following average:
\begin{eqnarray}
\nonumber
\left\langle \frac1{\xi_R}\right\rangle&=&\sum_x \left\langle p_R^2(x)
\right\rangle=\sum_x\frac{1}{R^2}\left\langle \sum_{j,l=1}^R
\delta_{xx_j}\,\delta_{xx_l}\right\rangle=\\
\nonumber
&=&\sum_x\left(\frac{1}{R}\,p(x)+\frac{R-1}{R}\,p^2(x)\right)\\
\label{app_eq4}
&=&\rho+(1-\rho)\frac1\xi\quad{\rm with}\quad \rho=\frac1R\ .
\end{eqnarray}
Here the first term arises from the $(j=l)$- and the second term from the 
$(j\neq l)$-contributions. Wee see that for a finite ratio $R/\xi$ the average 
histogram-IPR does not coincide with the exact IPR. Eq. (\ref{app_eq4}) 
allows for the numerical extrapolation:
\begin{equation}
\label{app_eq5}
\xi_\infty=\xi_R\,\frac{1-1/R}{1-\xi_R/R}=\xi_R\,\frac{1-\rho}{1-\rho\,\xi_R}
\end{equation}
where $\xi_R$ is the numerical histogram-IPR of which we hope that it is 
close to its average (for large enough $R$) thus justifying (\ref{app_eq5}). 
This extrapolated IPR will be more reliable than the histogram-IPR for 
moderate values of $R$ and allow for a more accurate determination of the 
functional dependence of the IPR on the different parameters. However, 
$\xi_\infty$ is still subject to statistical errors and therefore we also need 
to compute the variance of the histogram-IPR and related to this we also 
need the average of the second order histogram-IPR:
\begin{equation}
\label{app_eq6}
\frac{1}{\xi_{2,R}}=\sum_x\,p_R^3(x)
\end{equation}
as compared to the exact second order IPR: 
\begin{equation}
\label{app_eq7}
\frac{1}{\xi_2}=\sum_x\,p^3(x)\ .
\end{equation}
This quantity is comparable to $1/\xi^2$. Actually, using Cauchy-Schwartz 
inequality (for two vectors $v_x=p(x)^{1/2}$ and $w_x=p(x)^{3/2}$) we find:
\begin{equation}
\label{app_eq8}
\frac{1}{\xi_2}=\bra{v}\ket{v}\bra{w}\ket{w}
\ge |\bra{v}\ket{w}|^2 = \frac{1}{\xi^2}
\end{equation}
with the standard scalar product: $\bra{v}\ket{w}=\sum_x\,v_x\,w_w$. 
In Eq. (\ref{app_eq8}) we have equality if $p(x)=$const.$>0$ for certain 
values of $x$ and $p(x)=0$ for the other values of $x$.

Repeating the calculation (\ref{app_eq4}) for the second order IPR we find:
\begin{eqnarray}
\nonumber
\left\langle \frac1{\xi_{2,R}}\right\rangle&=&\sum_x \left\langle p_R^3(x)
\right\rangle=\frac{1}{R^3}\sum_x\left\langle \sum_{j,l,k=1}^R
\delta_{xx_j}\,\delta_{xx_l}\,\delta_{xx_k}
\right\rangle\\
\nonumber
&=&\frac{1}{R^3}\sum_x\Bigl(
R\,p(x)+3R(R-1)\,p^2(x)+\\
\nonumber
&&+R(R-1)(R-2)\,p^3(x)\Bigr)\\
\label{app_eq9}
&=&\rho^2+3\,\rho(1-\rho)\frac1\xi+
(1-\rho)(1-2\,\rho)\frac1\xi_{2}
\end{eqnarray}
where the three contributions correspond to the cases where the 
three values $j$, $l$ and $k$ are equal, or only two of them or none of them 
are equal. 

The evaluation of the variance of the histogram-IPR is more complicated 
but straight forward:
\begin{equation}
\label{app_eq10}
\left\langle \frac1{\xi_R^2}\right\rangle=\sum_x f_1(x)+\sum_{x\neq y} 
f_2(x,y)
\end{equation}
with:
\begin{eqnarray}
\nonumber
f_1(x)&=&\left\langle p_R^4(x)\right\rangle\\
\nonumber
&=&\rho^3\,p(x)+7\,\rho^2(1-\rho)\,p^2(x)+\\
\label{app_eq11}
&&+6\,\rho(1-\rho)(1-2\rho)\,p^3(x)+\\
\nonumber
&&+(1-\rho)(1-2\rho)(1-3\rho)\,p^4(x)\ .
\end{eqnarray}
We note that the prefactor ``7'' in the second term of (\ref{app_eq11}) 
arises from 
4 permutations of the type $j=l=k\neq m$ and 3 permutations 
of the type $j=l\neq k=m$ in the summation index. The prefactor ``6'' 
in the third term arises from 6 permutations of the type $j=l\neq k\neq m$.

Furthermore for $x\neq y$ we obtain:
\begin{eqnarray}
\nonumber
f_2(x,y)&=&\left\langle p_R^2(x)\,p_R^2(y)\right\rangle\\
\label{app_eq12}
&=&\rho^2(1-\rho)\,p(x)\,p(y)+\\
\nonumber
&&+\rho(1-\rho)(1-2\rho)\,\left[p^2(x)\,p(y)+p(x)\,p^2(y)\right]\\
\nonumber
&&+(1-\rho)(1-2\rho)(1-3\rho)\,p^2(x)\,p^2(y)
\end{eqnarray}
and therefore:
\begin{eqnarray}
\label{app_eq13}
\left\langle \frac1{\xi_R^2}\right\rangle-
\left\langle \frac1{\xi_R}\right\rangle^2&=&
2\,\rho^2(1-\rho)\left(\frac1\xi-\frac1{\xi^2}\right)+\\
\nonumber
&&+4\,\rho(1-\rho)(1-2\rho)\left(\frac{1}{\xi_2}-\frac1{\xi^2}\right)\ .
\end{eqnarray}
In the numerical scheme we determine $\xi_R^{-1}$ and $\xi_{2,R}^{-1}$ for 
one realization of $x_1,\,\ldots,\,x_R$ and using (\ref{app_eq4}), 
(\ref{app_eq9}) we determine approximate values of $\xi^{-1}$ 
and $\xi_2^{-1}$ and by (\ref{app_eq13}) the variance of $\xi_R^{-1}$. 
The number $R$ of $x_j$-values is increased until the relative error is below 
a certain threshold, typically 2\%. 
We note that according to (\ref{app_eq13}) for the special case 
$\xi_2=\xi^2$ the variance scales with $\rho^2=R^{-2}$ and not with the 
usual behavior $\rho=R^{-1}$. 
Even for $\xi_2<\xi^2$ the numerical prefactor of the $R^{-1}$ 
term may be quite suppressed as compared to the $R^{-2}$-term 
and therefore it is better to be careful and not to neglect this term in 
(\ref{app_eq13}).


\end{document}